\documentclass[twoside,leqno]{article}

\pdfoutput=1

\usepackage{RR}
\usepackage[a4paper]{geometry}
\usepackage{a4wide}

\usepackage{caption}
\usepackage[utf8]{inputenc}
\usepackage[OT1]{fontenc} 

\usepackage[english]{babel}

\usepackage[numbers]{natbib}

\usepackage[colorlinks=true, citecolor=blue, linkcolor=blue]{hyperref}

\usepackage{amssymb, amsmath, amsthm, mathtools}
\usepackage{amsfonts, bbold, dsfont}

\usepackage{color}
\definecolor{darkred}{rgb}{0.55, 0.0, 0.0}
\definecolor{darkgreen}{rgb}{0.0, 0.55, 0.13}
\definecolor{darkblue}{rgb}{0.0, 0.0, 0.55}
\definecolor{darkviolet}{rgb}{0.58, 0.0, 0.83}

\usepackage{listings, lstcoq}
\newcommand*{\code}[1]{\texttt{#1}}

\usepackage{tikz}
\usetikzlibrary{positioning,shapes,calc,trees,arrows,fit}
\usepackage{subcaption}

\newtheorem{Theorem}{Theorem}
\newtheorem{Definition}[Theorem]{Definition}
\newtheorem{Lemma}[Theorem]{Lemma}

\newcommand{\AuthorsList}{%
  S. Boldo, F. Clément, \& Louise Leclerc}

\newcommand{\LMF}{%
  Université Paris-Saclay, CNRS, ENS Paris-Saclay, Inria,
  Laboratoire Méthodes Formelles,
  91190, Gif-sur-Yvette, France.}
\newcommand{\SERENA}{%
  a. Inria, 2 rue Simone Iff, 75589 Paris, France.\protect\\
  b. CERMICS, École des Ponts, 77455 Marne-la-Vallée, France.}
\newcommand{\DMA}{%
  DMA, École normale supérieure, PSL University, CNRS, 75005 Paris, France.}

\newcommand{\Title}{%
  A Coq Formalization of the Bochner Integral}
\newcommand{\Titre}{%
  Une formalisation en Coq de l'intégrale de Bochner}

\newcommand{\Abstract}{%
  The Bochner integral is a generalization of the Lebesgue integral, for
  functions taking their values in a Banach space.
  Therefore, both its mathematical definition and its formalization in the Coq
  proof assistant are more challenging as we cannot rely on the properties of
  real numbers.
  Our contributions include an original formalization of simple functions,
  Bochner integrability defined by a dependent type, and the construction of
  the proof of the integrability of measurable functions under mild hypotheses
  (weak separability).
  Then, we define the Bochner integral and prove several theorems, including
  dominated convergence and the equivalence with an existing formalization of
  Lebesgue integral for {\nonnegative} functions.
}
\newcommand{\Resume}{%
  L'intégrale de Bochner est une généralisation de l'intégrale de Lebesgue pour
  des fonctions à valeurs dans un espace de Banach.
  Sa définition mathématique et sa formalisation dans l'assistant de preuve Coq
  en sont donc plus difficiles puisque l'on ne peut pas s'appuyer sur les
  propriétés des nombres réels.
  Nos contributions incluent une formalisation originale des fonctions simples,
  l'intégrabilité de Bochner définie par un type dépendant, et la construction
  de la preuve de l'intégrabilité de fonctions mesurables sous une hypothèse de
  séparabilité faible.
  Puis, nous définissons l'intégrale de Bochner et prouvons plusieurs
  théorèmes, dont la convergence dominée et l'équivalence avec une
  formalisation préexistante de l'intégrale de Lebesgue pour les fonctions
  mesurables positives.
}

\newcommand{\Keywords}{%
  Formal proof,
  Coq,
  Measure theory,
  Bochner integration
}
\newcommand{\Motscles}{%
  Preuve formelle,
  Coq,
  Théorie de la mesure,
  Intégrale de Bochner
}

\newcommand*{\eg}{e.g.}
\newcommand*{\ie}{i.e.}

\newcommand*{\nonempty}{non\-empty}
\newcommand*{\nonnegative}{non\-negative}

\newcommand*{\Record}{record} 

\newcommand*{\soft}[1]{\textsf{#1}} 
\newcommand*{\Coq}{\soft{Coq}}
\newcommand*{\Coquelicot}{\soft{Coquelicot}}
\newcommand*{\HOLfour}{\soft{HOL4}}
\newcommand*{\IsabelleHOL}{\soft{Isabelle/HOL}}
\newcommand*{\Lean}{\soft{Lean}}
\newcommand*{\Mizar}{\soft{Mizar}}
\newcommand*{\PVS}{\soft{PVS}}

\newcommand*{\IsabelleAnalysisLib}{%
  https://isabelle.in.tum.de/dist/library/HOL/HOL-Analysis}
\newcommand*{\LeanMathLib}{https://leanprover-community.github.io/mathlib_docs}
\newcommand*{\PVSLib}{%
  https://shemesh.larc.nasa.gov/fm/ftp/larc/PVS-library/library}

\newcommand*{\intInter}[2]
{\ensuremath{%
  \left[\!\left[ \ensuremath{{#1}},\, \ensuremath{{#2}} \right]\!\right]
}}
\newcommand*{\norm}[1]{\|#1\|}
\newcommand*{\eps}{\varepsilon}

\newcommand{\calSF}{\mathcal{SF}}
\newcommand{\calSFplus}{\calSF_+}
\newcommand{\intSFplus}{\int_{\calSFplus}}
\newcommand{\calM}{\mathcal{M}}
\newcommand{\calMplus}{\calM_+}
\newcommand{\intMplus}{\int_{\calMplus}}

\newcommand*{\N}{\mathbb{N}}
\newcommand*{\Q}{\mathbb{Q}}
\newcommand*{\R}{\mathbb{R}}
\newcommand*{\Rbar}{\overline{\R}}
\newcommand*{\C}{\mathbb{C}}

\newcommand{\coeur}[1][fill=white]{%
  \draw[#1] (0,0)
  .. controls +(0,2) and +(0,2) .. (3,0)
  .. controls +(0,-2) and +(0,2) .. (0,-4)
  .. controls +(0,2) and +(0,-2) .. (-3,0)
  .. controls +(0,2) and +(0,2) .. (0,0);
}

\newcommand{\oftype}[2]{$#1$ of type $#2$}

\RRdate{February 2022}

\RRauthor{%
  Sylvie Boldo\thanks{{\LMF} \texttt{sylvie.boldo@inria.fr}}
  \and François Clément\thanks{{\SERENA} \texttt{francois.clement@inria.fr}}
  \and Louise Leclerc\thanks{{\DMA} \texttt{louise.leclerc@ens.psl.eu}}
}
\authorhead{\AuthorsList}

\RRtitle{\Titre}
\RRetitle{\Title}
\titlehead{\Title}

\RRresume{\Resume}
\RRabstract{\Abstract}

\RRmotcle{\Motscles}
\RRkeyword{\Keywords}

\RRprojets{Toccata and Serena}

\RCSaclay

\begin{document}

\RRNo{9456}
\makeRR

\section{Introduction}

This work is devoted to the {\Coq} formalization of the Bochner integral.
Among a huge variety of integrals, {\eg} see~\cite{bur:gi:07}, the Bochner
integral~\cite{boc:ifw:33} is a generalization of the Lebesgue integral, for
real-valued functions, to the case of functions taking their values in a Banach
space, {\ie} a complete normed vector space.  Thus, it is perfectly suited for
the study of partial differential equations involving time and space variables.
For instance, given a real number~$T>0$ and a regular enough space domain
$\Omega\subset\R^3$, one might be interested in integrating functions mapping
the time interval~$[0,T]$ to the Hilbert space~$L^2(\Omega)$ of
functions~$\Omega\to\R$ that are square Lebesgue-integrable, which of course is
also a Banach space.  In a formal proof setting, it also allows us to have a
single definition and set of theorems for integrating on either~$\R$, $\C$,
or~$\R^n$.

The building of the Bochner integral follows a similar scheme to that of the
Lebesgue integral: first consider simple functions, that only take a finite
number of values, define their integral by summing terms of the form
\emph{measure of preimage $\times$ value}, and then extend to the limit of
simple functions.  The main difference here is the absence of order in a normed
vector space, which prevents the use of monotonicity and of the LUB property,
as in~$\R$, and thus prohibits infinite terms in the integral.  Instead, it
relies on completeness, and on the additional assumption of separability of the
Banach space, or at least of the range of the integrand function.  Note that
some mathematical authors and the other formalizations prefer second
countability, which is stronger than separability in general, but actually
equivalent in the case of metric spaces (and Banach spaces are).  Rather than
the seminal paper by S.~Bochner, or the monograph by
J.~Mikusi\'nski~\cite{mik:bi:78}, we chose to follow the modern presentation of
the course in real analysis by G.~Teschl~\cite{tes:tra:21}.

\medskip

The formalization is available at the following link:
\begin{center}
  {\small \url{https://lipn.univ-paris13.fr/coq-num-analysis/tree/Bochner.1.0/Lebesgue/bochner_integral}}\\
  where the tag \texttt{Bochner.1.0} corresponds to the code of this article.
\end{center}

\medskip

The paper is organized as follows.  After drawing up the state of the art in
Section~\ref{sec:soa}, Section~\ref{sec:Lebesgue} presents the {\Coq}
formalization of the Lebesgue integral~\cite{BCF21} we rely on, and
Section~\ref{sec:topo} some preliminary topological results.
Section~\ref{sec:simple} is dedicated to simple functions.  Bochner
integrability is addressed in Section~\ref{sec:Bochner-integrability}, and the
Bochner integral is defined in Section~\ref{sec:integral}.  Finally,
Section~\ref{sec:concl} concludes and gives some perspectives.

\section{State of the art}
\label{sec:soa}

Measure theory and {\nonnegative} Lebesgue integration has been formalized in
formal proof assistants such as {\Mizar}\footnote{\url{https://fm.mizar.org/}},
{\PVS}~\cite{PVS_Ref}, {\IsabelleHOL}~\cite{Isabelle_Ref},
{\HOLfour}\footnote{\url{https://hol-theorem-prover.org/}},
{\Lean}~\cite{Lean_Ref,Lean_Maths}, and
{\Coq}\footnote{\url{https://coq.inria.fr/refman/}}.  We may
cite~\cite{Bialas1991,Endou2017} in {\Mizar}, dedicated libraries in
{\PVS}\footnote{%
  \url{\PVSLib/measure_integration.html}, \url{\PVSLib/lebesgue.html}},
{\IsabelleHOL}\footnote{%
  \url{\IsabelleAnalysisLib/Measure_Space.html},
  \url{\IsabelleAnalysisLib/Nonnegative_Lebesgue_Integration.html}}, and
{\Lean}\footnote{%
  \url{\LeanMathLib/measure_theory/measure/measure_space.html},
  \url{\LeanMathLib/measure_theory/integral/lebesgue.html}},
\cite{MhamdiHT10} in {\HOLfour}, and dedicated libraries\footnote{%
  \url{https://github.com/math-comp/analysis/}} and~\cite{BCF21} in {\Coq}.

\medskip

There are few proof assistants that provide the Bochner integral, while the
Riemann or Lebesgue integrals are more widespread.  To the best of our
knowledge, there are already two available formalizations.

First, {\IsabelleHOL} provides Bochner integrability and integral and the
dominated convergence theorem~\cite{AviHolSer17}.  Their goal is
probability and the central limit theorem and this generic integral easily
encompasses $\C$ and $\R^n$.  They assume a second-countable topology (while we
add the weaker separability hypothesis at the only needed point).  Their
definitions are rather similar to ours except for Bochner integrability: $f$ is
Bochner-integrable if and only if $f$ is measurable and its $L^1$-norm is
finite (which is equivalent to saying that $f$ is absolutely integrable).  See
our definition in Section~\ref{sec:Bochner-integrability}.

Second and last, {\Lean} provides Bochner integrability and integral, the
dominated convergence theorem and the Fubini theorem~\cite{VanDoo21}.  They
assume second-countable real Banach space.  The main difference is that the
quotient space~$L^1$ is defined and that the Bochner integral applies to
equivalent classes of functions.  Moreover, the definition of the integral is
very different from ours: they extend to~$L^1$ the continuous linear map that
is the integral on integrable simple functions.

As a conclusion on this state of the art, the proved theorems are similar,
contrary to the definitions.  A difference is that they require
second-countability while we locally require weak separability.  Another
difference is the simple function definitions: they both rely on the fact that
the image is a finite set while we use a dependent type.  See our definition in
Section~\ref{sec:simple}.

\medskip

The present formalization uses the real standard library of {\Coq}, based
on~\cite{May01}, and the {\Coquelicot} library~\cite{BLM15} extension.
These libraries provide support for classical real numbers, which is consistent
with the fact that the mathematics we are formalizing are based on classical
logic, as most of the real analysis results do.

\section{The Lebesgue integral in {\Coq}}
\label{sec:Lebesgue}

This work is based on several existing {\Coq} libraries.  We of course rely on
the standard library, and in particular for the real numbers~\cite{May01}.

We also rely on {\Coquelicot}~\cite{BLM15}, which is a conservative extension
of the real numbers.  We use several features of this library: the extended
real numbers and their operations; the algebraic hierarchy in particular for
the normed modules and Banach spaces; the underlying topology based on filters.
We refer the reader to \cite{BLM15} for more details.

We have also taken inspiration from a recent work defining the Lebesgue
integral for {\nonnegative} functions~\cite{BCF21}%
\footnote{\url{https://lipn.univ-paris13.fr/MILC/}} and require its latest
version.  Here are the important design choices of this library.  The
measurability of subsets of~$X$ is formalized as an inductive type
parameterized by \coqe{gen : (X -> Prop) -> Prop}, that represents the
corresponding generated $\sigma$-algebra.  When~$X$ is a metric space, in
{\Coquelicot} \coqe{X : UniformSpace}, one generally uses the Borel
$\sigma$-algebra that is generated by all the open subsets.

Simple functions are based on \emph{lists}.  More precisely, a function is a
simple function when its image is included in a finite list of values.  Then
this list may be canonized (by removing unused values and duplicates and
sorting the values) and this canonical list is used to compute the integral of
a simple function (provided a given measure).  This cannot be applied here as
Banach space values cannot be sorted, contrary to real numbers.

The integral for {\nonnegative} measurable functions is then defined as in
mathematics textbooks:
\[
  \intMplus f \, d\mu
  \, = \,
  \sup_{\substack{\psi \in \calSFplus\\\psi \leq f}} \intSFplus \psi \, d\mu
\]
with $\calSFplus$ being the set of {\nonnegative} simple functions, and
$\calMplus$ the set of {\nonnegative} measurable functions.  The integral of a
{\nonnegative} measurable function $f$ is the supremum of the integral of the
{\nonnegative} measurable simple functions~$\psi$ less than or equal to~$f$
pointwise.  Basic lemmas (such as monotony, scalar multiplication, addition)
are provided in \cite{BCF21}, as well the Beppo Levi (monotone convergence)
theorem and Fatou's lemma.

\medskip

For the sake of readability in the sequel, we do not always specify the scope
in the {\Coq} scripts.

\section{Some topology in normed modules}
\label{sec:topo}

We present here some preliminary needed results: a few lemmas above
{\Coquelicot} are given in Section~\ref{sec:lim_seq} and separability is
described in Section~\ref{sec:separability}.

\subsection{Additions to {\Coquelicot}}
\label{sec:lim_seq}

In order to formalize the Bochner integral in {\Coq}, one needs at first some
topology in normed vector spaces, and especially in Banach spaces.  In this
development, we choose to use the existing formalization of filters and open
subsets of {\Coquelicot}~\cite{BLM15}.  The main notion required is the limit
of sequences, which is straightforwardly given within {\Coquelicot}.  Given a
type \coqe{S : UniformSpace}, we may denote
\begin{lstlisting}
Definition lim_seq (u : nat -> S) := lim (filtermap u eventually).
\end{lstlisting}

Starting from this definition, one can easily prove its equivalence with the
more common textbook definition, that may also be more practical than filters
in some cases.
\begin{lstlisting}
Lemma is_lim_seq_epsilon {A : AbsRing} {E : NormedModule A} :
  \forall u : nat -> E, \forall l : E, is_lim_seq u l <->
    \forall \eps, 0 < \eps -> \exists N, \forall n, N \le n -> \| minus (u n) l \| < \eps.
\end{lstlisting}
Another useful lemma we may derive from this notion is the following, stating
the (Borel) measurability of a pointwise limit of measurable functions in any
vector space.  (Actually, this is not as easy as in the real case, where one
may use \coqe{LimSup} and \coqe{LimInf} to get a simple proof.)
\begin{lstlisting}
Lemma measurable_fun_lim_seq {X : Set} {gen : (X -> Prop) -> Prop} :
  \forall s : nat -> X -> E, (\forall n, measurable_fun gen open (s n)) ->
    \forall f : X -> E, (\forall x : X, is_lim_seq (\fun n => s n x) (f x)) -> measurable_fun gen open f.
\end{lstlisting}

\medskip

Similarly, we define equivalent Cauchy sequences in a normed vector space,
which may be easier to handle than Cauchy filters in practice:
\begin{lstlisting}
Definition NM_Cauchy_seq {A : AbsRing} {E : NormedModule A} (u : nat -> E) : Prop :=
  \forall \eps, \eps > 0 -> \exists n, \forall p q, p \ge n -> q \ge n -> ball_norm (u p) \eps (u q).
\end{lstlisting}

\subsection{Separability}
\label{sec:separability}

\begin{figure}[t]
  \centering
  \mbox{}
  \hfill
  \begin{subfigure}[t]{0.45\textwidth}
    \centering
    \begin{tikzpicture}[%
      ele/.style={fill=black,circle,minimum width=.8pt,inner sep=1pt},
      every fit/.style={ellipse,draw,inner sep=-2pt}]

      \draw (0,0) grid[step=0.5] (6,3);
      \foreach \k in {0,0.5,...,5.5}
        \foreach \i in {0,0.5,...,2.5}
          {\node[ele] at (\k+.25,\i+0.25) {};}
    \end{tikzpicture}
    \label{fig:sep1}
    \caption{The full space (in 2D) is separable with the given points.}
  \end{subfigure}
  \hfill
  \begin{subfigure}[t]{0.45\textwidth}
    \centering
    \begin{tikzpicture}[%
      ele/.style={fill=black,circle,minimum width=.8pt,inner sep=1pt},
      every fit/.style={ellipse,draw,inner sep=-2pt}]

      \draw (0,0) grid[step=0.5] (6,3);
      \foreach \k in {0,0.5,...,5.5}
        \foreach \i in {0,0.5,...,2.5}
          {\node[ele] at (\k+.25,\i+0.25) {};}
      \coeur[shift={(3,2)},scale=0.38,very thick,fill=red,opacity=0.3]
    \end{tikzpicture}
    \label{fig:sep2}
    \caption{The subset $Y$ is heart-shaped. For ensuring its separability,
      we may provide points inside~$Y$. But it is easier to only ensure
      weak separability by considering the same points as on the left, that may
      or may not belong to~$Y$.}
  \end{subfigure}
  \hfill
  \mbox{}
  \caption{Separability and weak separability: a figurative view.}
  \label{fig:sep}
\end{figure}
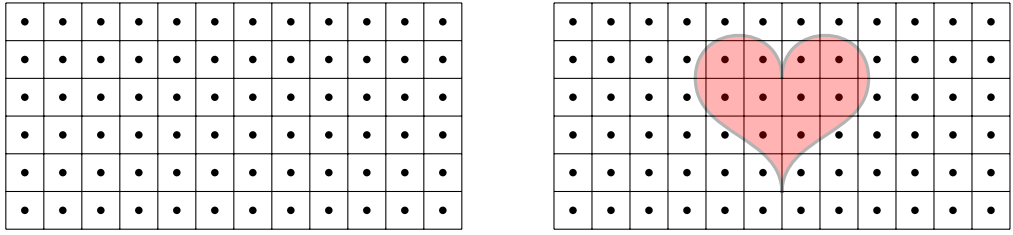

Next, we need a formalization of separability in normed vector spaces.  Let us
remind the mathematical definition of this property.

\begin{Definition}[separability]
  A topological space $(E,\,\tau)$ is said \emph{separable} when it contains a
  countable dense subset, {\ie} when there exists a sequence
  $(u_n)_{n\in\N}\in E^\N$ such that ($U$ is any {\nonempty} open subset)
  $$\forall\, U \in \tau,\;
  U \neq \emptyset \Rightarrow
  U \cap \{u_n \,|\, n \in \N \} \neq \emptyset.$$
\end{Definition}

It suffices in our case to define a more practical weaker version in which we
do not require the countable part to dwell inside the separable one.
\begin{Definition}[weak separability]
  Let $(E,\,\tau)$ be a topological space.
  A subset $Y\subseteq E$ is said \emph{weakly separable in~$E$} when
  there exists a sequence $(u_n)_{n\in\N}\in E^\N$ such that
  $$\forall\, U \in \tau,\;
  U \cap Y \neq \emptyset \Rightarrow
  U \cap \{u_n \,|\, n \in \N \} \neq \emptyset.$$
\end{Definition}

For example, let $E$ be $\R$ equipped with the usual topology, $Y$ be
$\R\setminus\Q \subseteq E$ and $(u_n)_{n\in\N}$ be a sequence whose range is
exactly $\Q$.  Then $Y$ (with the induced topology) and $(u_n)_{n\in\N}$ does
not satisfy the first definition since $(u_n)_{n\in\N}$ is not a sequence in
$Y$, but we may say from the second definition that $Y$ is weakly separable in
$E$, thus avoiding the building of a sequence of irrationals in~$Y$.  A more
visual and figurative example is given in Figure~\ref{fig:sep}.

For normed modules, the norm induces the topology, therefore we get the
following characterization, easier to formalize.
\begin{Lemma}[weak separability in normed vector spaces]
  Let $(E,\,\norm{\cdot})$ be a normed vector space.
  A subset $Y\subseteq E$ is weakly separable in $E$ if and only if there
  exists a sequence $(u_n)_{n\in\N}\in E^\N$ such that
  $$\forall y \in Y,\; \forall \eps > 0,\; \exists n \in \N,\;
  \norm{y - u_n} < \eps.$$
\end{Lemma}

Given \coqe{E:NormedModule R_AbsRing}, we define it in {\Coq} as
\begin{lstlisting}
Definition NM_seq_separable_weak (u : nat -> E) (P : E -> Prop) : Prop :=
  \forall x : E, P x -> \forall \eps : posreal, \exists n, ball_norm x \eps (u n).
\end{lstlisting}
Note that the sequence $u$ is explicit in this definition.

For instance, {\Coq} real numbers are weak separable.
\begin{lstlisting}
Lemma NM_seq_separable_weakR :
  NM_seq_separable_weak (\fun n => Q2R (bij_NQ n)) (\fun _ : R_NormedModule => True).
\end{lstlisting}
The sequence $u$ ranges over the rationals, relying on the bijection
\coqe{bij_NQ} from $\N$ onto $\Q$.

\section{Formalizing simple functions}
\label{sec:simple}

A first important step towards Bochner integrability and integral is the
definition of simple functions on a Banach space.  Even if the Lebesgue
integral also needs simple functions~\cite{BCF21}, their formalization is not
applicable in our case and we have provided an original definition described in
Section~\ref{sec:simple_def}, as well as the Bochner-integrability.  The value
of the integral is given in Section~\ref{sec:simple_int}.

\subsection{Definition and properties}
\label{sec:simple_def}

Following~\cite{tes:tra:21}, we start by formalizing simple functions.  Then
in Section~\ref{sec:integral}, we define the Bochner integral as a limit
of integrals of simple functions (as for the Lebesgue integral).

Let us consider a measurable space $(X,\Sigma)$, and~$E$ a normed vector space
that is assumed to be equipped with its Borel $\sigma$-algebra (generated by
all open subsets).  In {\Coq}, we have \coqe{X : Set}, the
$\sigma$-algebra~$\Sigma$ is represented by some generator
\coqe{gen : (X -> Prop) -> Prop} (see Section~\ref{sec:Lebesgue}),
\coqe{E : NormedModule A} with \coqe{A : AbsRing}, and its Borel
$\sigma$-algebra is generated by the generic \coqe{open : (E -> Prop) -> Prop}.
Then, the mathematical definition of (measurable) simple function is the
following.

\begin{Definition}[simple function]\label{def:sf1}
  A function $f : X \rightarrow E$ is said \emph{simple} when its range is
  finite and all the preimages are measurable.
\end{Definition}

In~\cite{BCF21}, as explained in Section~\ref{sec:Lebesgue}, the simple
functions for the Lebesgue integral were defined by the existence of a list
that collects the values taken by the function.  This was chosen because by
forcing the list to be sorted, and not to contain any duplicates or unnecessary
value, one gets a canonical representation of a simple function.  However, this
is no longer possible with vector-valued functions, where no order can be used
on the image space.  But it is known that Definition~\ref{def:sf1} is
equivalent to the two following
characterizations.

\begin{Lemma}[characterization 1]
  A function $f : X \rightarrow E$ is simple if and only if it is a linear
  combination of characteristic functions of measurable subsets.
\end{Lemma}

\begin{Lemma}[characterization 2] \label{lem:charac2}
  A function $f : X \rightarrow E$ is simple if and only if there exists a
  finite partition $(A_i)_{i\in I}$ of\/ $X$ such that for all $i\in I$, $f$ is
  constant over~$A_i$, and $A_i$ is measurable.
\end{Lemma}

\begin{figure}[t]
  \centering
  \begin{tikzpicture}[%
    ele/.style={fill=black, circle, minimum width=.8pt, inner sep=1pt},
    every fit/.style={ellipse, draw, inner sep=-2pt},
    mapsto/.style={->, thick, shorten >=2pt, shorten <=2pt, >=stealth}]

    \node (f) at (-1.5,6.3) {$f\,:$};

    \node (X) at (-1,6.3) {$X$};

    \node[ele] (x0) at (-1,5) {};
    \node[ele] (x1) at (-1,4) {};
    \node[ele] (x2) at (-1,3) {};
    \node[ele] (x3) at (-1,2) {};
    \node[ele] (x5) at (-1,1) {};
    \node[ele,label=left:$x$] (x4) at (-1,0) {};

    \node (N) at (4,6.3) {$\N$};

    \node[ele, label=above:$0$] (n0) at (4,4.5) {};
    \node[ele, label=above:$1$] (n1) at (4,3.5) {};
    \node[ele, label=above:$2$] (n2) at (4,2.5) {};
    \node[ele, label=above:\coqe{which}$(x)$] (n3) at (4,1.5) {};
    \node[ele, label=below:\coqe{max_which}] (n4) at (4,0.5) {};

    \node (E) at (10,6.3) {$E$};

    \node[ele, label=right:\coqe{val}$(0)$] (e0) at (10,4) {};
    \node[ele, label=right:\mbox{\coqe{val}$(1)$ = \coqe{val}$(2)$}] (e1) at (10,3) {};
    \node[ele, label=right:\coqe{val (which}$(x)$\coqe{)}] (e2) at (10,2) {};
    \node[ele, label=right:$0$] (e3) at (10,1) {};

    \node[draw, fit=(x0) (x1) (x2) (x3) (x4) (x5), minimum width=2cm] {} ;
    \node[draw, fit=(n0) (n1) (n2) (n3) (n4), minimum width=3cm] {} ;
    \node[draw, fit=(e0) (e1) (e2) (e3), minimum width=5cm] {} ;

    \draw[mapsto] (x1) -- (n4);
    \draw[mapsto] (x2) -- (n1);
    \draw[mapsto] (x3) -- (n0);
    \draw[mapsto, dotted] (x4) -- (n3);
    \draw[mapsto] (x0) -- (n1);
    \draw[mapsto] (x5) -- (n2);

    \draw[mapsto] (X) -- (N) node [midway, above] {\coqe{which}};

    \draw[mapsto] (n0) -- (e0);
    \draw[mapsto] (n1) -- (e1);
    \draw[mapsto, dotted] (n3) -- (e2);
    \draw[mapsto] (n4) -- (e3);
    \draw[mapsto] (n2) -- (e1);

    \draw[mapsto] (N) -- (E) node [midway, above] {\coqe{val}};
  \end{tikzpicture}
  \caption{%
    {\Coq} representation of simple functions.\\
    Given a cutting of the set~\coqe{X} into~\coqe{max_which+1} pairwise
    disjoint measurable parts, the function \coqe{which : X -> nat} maps
    elements of each part to a distinct index, an integer in the range
    $\intInter{0}{\mbox{\coqe{max_which}}}$.
    Then, the function \coqe{val : nat -> E} maps each index of the previous
    range to some vector value in~\coqe{E}.
    The greatest index \coqe{max_which} is mapped to zero.
    The represented simple function $f:X\to E$ is actually the composition of
    \coqe{which} and \coqe{val}.
    Note that several indices (here~\coqe{1} and~\coqe{2}) may be mapped to
    the same vector value~$v$ (here \coqe{val 1} equals \coqe{val 2}), meaning
    that the preimage $f^{-1}(\{v\})$ is actually the (disjoint) union of the
    parts mapped to the indices.
    Thus, the representation is not unique.
    Of course, parts need not be convex, nor connected, and some of them may be
    empty, including the last one, associated with the value zero.}
  \label{fig:simple_fun}
\end{figure}

We tried to formalize both in {\Coq} and the second proved to be much more
efficient to handle.

The chosen data structure takes the following form.
\begin{lstlisting}
Record simpl_fun := mk_simpl_fun {
  which : X -> nat;
  val : nat -> E;
  max_which : nat;
  ax_val_max_which : val max_which = zero;
  ax_which_max_which : \forall x : X, which x \le max_which;
  ax_measurable : \forall n : nat, n \le max_which -> measurable gen (\fun x => which x = n);
}.
\end{lstlisting}

Such a {\Record} tells us that in order to build a simple function, there are
three values that should be given to {\Coq}: \coqe{which}, \coqe{val}, and
\coqe{max_which}, see Figure~\ref{fig:simple_fun}, and several proofs.  The
function \coqe{which} corresponds to a cutting of the space $X$, or to an index
in the finite set of preimages.  The integer \coqe{max_which} is the maximal
value allowed for \coqe{which} (it ensures the finiteness of the cutting).  The
function \coqe{val} provides the value corresponding to a given integer.

For instance, suppose we want to construct in {\Coq} the simple function
corresponding to \oftype{f}{X \rightarrow E} with the finite partition
$(A_i)_{i\in I}$ of Lemma~\ref{lem:charac2}.  First of all, because $I$ is
finite, we can suppose that it is of the form $\intInter{0}{n}$ for some
$n\in \N$ (implicitly, here we also suppose that $X$ is not empty), then:
\begin{itemize}
\item
  the number of parts minus one in our cutting ({\ie} $|I|-1$ or $n$ in the
  above description) is stored into \coqe{max_which};
\item
  because $(A_i)_{i\in I}$ forms a pairwise disjoint cover of~$X$, for each
  $x\in X$, there exists a unique $i\in I$ such that $x\in A_i$.  The function
  \coqe{which} associates this $i\in \N$ with each $x\in X$.  So inside
  {\Coq}, this becomes \coqe{which : X -> nat}.
\item
  Finally, for every $i\in I$, $f$ takes over $A_i$ a value $v_i \in E$.  This
  is stored inside \coqe{val i}, for every \coqe{i : nat}.  As we see below,
  the value of \coqe{val} for $i > n$ does not matter in our formalization.
\end{itemize}

In addition to these three values, there is a need for properties that ensure
such a structure correctly represents a simple function and behaves nicely.
\begin{itemize}
\item
  First, we need to ensure that \coqe{val} takes the value \coqe{zero} on
  \coqe{max_which}.  The proof is stored into
  \coqe{ax_val_max}\linebreak[0]\coqe{_which}.
  This is not a mathematical consideration but a commodity in order to
  manipulate integrability of simple functions.  This allows us to deal
  with the preimage of $0$ separately from the others, and especially to allow
  $A_{n}$ to be of infinite measure for integrable simple functions.
\item
  Secondly, because \coqe{which} has its values in \coqe{nat} while
  mathematically, it should have it in $\intInter{0}{\mbox{\coqe{max_which}}}$,
  we must ensure that \coqe{which} does not exceed \coqe{max_which}.  The proof
  is stored into \coqe{ax_which_max_which}.
\item
  Finally, we have to ensure the measurability of the simple function, {\ie}
  the measurability of all its preimages.  The proof is stored into
  \coqe{ax_measurable}.
\end{itemize}

It is convenient to use the {\Record} defining a simple function as a function,
so we also define the following coercion.
\begin{lstlisting}
Definition fun_sf (sf : simpl_fun) : X -> E := \fun x => sf.val (sf.which x).

(* So we may write "sf x" for sf : simpl_fun E gen, and x : X. *)
Coercion fun_sf : simpl_fun \coerc Funclass.
\end{lstlisting}

For instance, for the indicator function of a (measurable) subset $A$, we
would have \coqe{which} that returns 0 on $A$ and 1 on $\neg A$;
\coqe{max_which} that is 1; and \coqe{val}$(n)$ that is 1 when $n=0$ and 0
elsewhere.  All assumptions hold.  Therefore, for $x\in A$, we
have \coqe{sf x = sf.val (sf.which x) = sf.val 0 = 1}, and for
$x\in \neg A$, we have \coqe{sf x = sf.val (sf.which x) = sf.val 1 = 0}.

Note that the type \coqe{simpl_fun} actually carries more structure than just
the definition of simple functions.  As a consequence, the representation of a
simple function~$f$ by an instance of the {\Record} is not unique.  Indeed, the
same value $v\in E$ could be associated with several distinct indices, meaning
that the actual preimage $f^{-1}(\{v\})$ could be represented by several
(pairwise disjoint) parts with distinct indices.  And of course, this may occur
for the value~\coqe{zero}, already associated with the index \coqe{max_which}.

Nevertheless, we may recover usual properties about simple functions.  An
interesting one is their measurability as defined in \cite{BCF21}.
\begin{lstlisting}
Lemma measurable_fun_sf : \forall sf : simpl_fun E gen, measurable_fun gen open sf.
\end{lstlisting}

\medskip

We have also explicitly constructed an instance of \coqe{simpl_fun E gen} for
the sum, opposite, subtraction, scalar product, norm or power of simple
functions.

As an example, given two simple functions $f,\, g : X \rightarrow E$ with
respective decomposition $(A_i)_{i\in\intInter{0}{n}}$ and
$(B_j)_{j\in\intInter{0}{m}}$, we get a correct decomposition for $f + g$ with
$(A_i \cap B_j)_{(i,\,j)\in\intInter{0}{n}\times\intInter{0}{m}}$.  To
formalize this decomposition in {\Coq}, we used an explicit bijection between
$\intInter{0}{n}\times\intInter{0}{m}$ and $\intInter{0}{(n+1)\cdot(m+1) - 1}$.

We use in the sequel the following notations:
\begin{lstlisting}
"sf + sg" := sf_plus sf sg.
"- sg" := sf_scal (opp one) sg.
"sf - sg" := sf_plus sf (sf_scal (opp one) sg).
"a \cdot sf" := sf_scal a sf.
"\| sf \|" := sf_norm sf.
"sf ^ p" := sf_power sf p.
\end{lstlisting}

\medskip

Over these simple functions, one also needs to define the integrability
property stating that preimages have a finite measure, except possibly for that
of zero (corresponding at least to the index \coqe{max_which}).  This allows
in Section~\ref{sec:simple_int} to sum terms of the form \emph{measure of
  preimage $\times$ value}.  For the sake of smoothness, we require all the
parts of index smaller than \coqe{max_which} to have finite measure.  It
therefore prevents parts of infinite measure with \coqe{val n = zero} and
\coqe{n < max_which}.  This is allowed in mathematics but impractical in formal
proofs and moreover, this case may be kept out (see below).
\begin{lstlisting}
Definition integrable_sf (sf : simpl_fun) :=
  \forall n, n < sf.max_which -> is_finite (\mu (\fun x => sf.which x = n)).
\end{lstlisting}

Indeed, this definition is not equivalent to the usual mathematical definition
of integrability.  Simple functions whose representation involves zero values
for indices \coqe{n < max_which} are not recognized as integrable when the
corresponding preimages have infinite measure.  But, both definitions match
when we ensure that the only part associated with the value zero is the last
one (with index \coqe{max_which}).  And this can be proved through the
following result.
\begin{lstlisting}
Lemma sf_remove_zeros (sf : simpl_fun E gen) :
  { sf' : simpl_fun E gen | (\forall x : X, sf x = sf' x) /\ (\forall n, n < sf'.max_which -> sf'.val n \neq zero) }.
\end{lstlisting}
Here, we used a \coqe{sig} from {\Coq}, that is a dependent type containing
an instance of \coqe{simple_fun E gen} together with a proof of
\coqe{(\forall x : X, sf x = sf' x)} and
\coqe{(\forall n, n < sf'.max_which -> sf'.val n \neq zero)}.  So we may use it
to remove unwanted zero values from our structure representing a simple
function.

The proof of this proposition is straightforward though tedious: browsing all
the values in \coqe{val}, suppressing the redundant zeros and redefining
\coqe{which} in order to have
\coqe{which x} = \coqe{sf'.max_which} each time we have that
\coqe{sf.val (sf.which x)} = \coqe{zero}.

Note also that the previous decomposition for the sum of simple functions
maintains the integrability.

\subsection{The Bochner integral for simple functions}
\label{sec:simple_int}

Now that we have defined simple functions, we are able to define the integral
of such functions.

Following Section~\ref{sec:simple_def}, let us now consider a measure space
$(X,\Sigma,\mu)$ where~$\mu$ is a measure on the measurable space $(X,\Sigma)$,
and~$E$ is now a normed vector space \emph{over~$\R$}.
In {\Coq}, we have now \coqe{\mu : measure gen}
and \coqe{E : NormedModule R_AbsRing}.
Then, we stick to the following mathematical definition.
\begin{Definition}[Bochner integral of simple function]
  Given an integrable simple function \oftype{s}{X\rightarrow E}, its \emph{Bochner
    integral (relatively to measure $\mu$ on $X$)} is defined by
  $$\int s \,d\mu := \sum_{v\in E} \mu\left(f^{-1}\{v\}\right) \cdot v,$$
  with the convention $\infty \cdot 0_E := 0_E$.
\end{Definition}
The former sum is finite according to the definition of integrable simple
function, and this may be translated in our formalization by
\begin{lstlisting}
Definition BInt_sf (\mu : measure gen) (sf : simpl_fun _ gen) : E :=
  sum_n (\fun n => scal (real (\mu (nth_carrier sf n))) (sf.val n)) (sf.max_which).
\end{lstlisting}
\noindent
where \coqe{nth_carrier} is the preimage defined by
\begin{lstlisting}
Definition nth_carrier (sf : simpl_fun) (n : nat) : (X -> Prop) := \fun x => sf.which x = n.
\end{lstlisting}

From this definition we may derive the usual properties of the integral such as
linearity,
\begin{lstlisting}
Lemma BInt_sf_lin {sf sg : simpl_fun E gen} (a b : R) :
  integrable_sf \mu sf -> integrable_sf \mu sg ->
  BInt_sf \mu (a \cdot sf + b \cdot sg) = a \cdot (BInt_sf \mu sf) + b \cdot (BInt_sf \mu sg).
\end{lstlisting}
Note that $\cdot$ is the scalar multiplication in the normed vector space.  The
mathematical proof implies some factorizations and finite sums inversion.  This
is basic linear algebra, but has proved slightly tedious.  Here, the main
difficulty is to handle the measure that takes values in $\Rbar$, and to manage
separately:
\begin{itemize}
\item
  $\alpha \cdot v$ when $v\in E$ and $\alpha \in \R$
  ({\ie} with {\Coquelicot} formalism, \coqe{is_finite \alpha});
\item
  $+ \infty \cdot 0_E$, which equals $0_E$ by mathematical and
  {\Coquelicot} conventions.
\end{itemize}

Another usual property is the triangle inequality.
\begin{lstlisting}
Lemma norm_Bint_sf_le (sf : simpl_fun E gen) : \| BInt_sf \mu sf \| \le BInt_sf \mu \| sf \|.
\end{lstlisting}
This lemma reduces by definition to the usual triangle inequality for a finite
sum.

\medskip

As explained, our formalization of simple function is not canonical, so it must
be proved that the value of \coqe{BInt_sf sf} only depends on the values taken
by \coqe{sf} and not on the cutting we chose to represent this function.  This
is stated in the following extensionality lemma:
\begin{lstlisting}
Lemma BInt_sf_ext {sf sf' : simpl_fun E gen} :
  integrable_sf \mu sf -> integrable_sf \mu sf' ->
  (\forall x : X, sf x = sf' x) -> BInt_sf \mu sf = BInt_sf \mu sf'.
\end{lstlisting}

\section{Bochner-integrable functions}
\label{sec:Bochner-integrability}

Now we define the integrability of functions, by the means of an approximation
by simple functions.

Following Section~\ref{sec:simple_int}, let us still consider a measure space
$(X,\Sigma,\mu)$, and~$E$ is now a \emph{Banach space} over~$\R$.
In {\Coq}, this becomes \coqe{E : CompleteNormedModule R_AbsRing}.
As in textbooks, we consider $f : X \rightarrow E$ as the pointwise limit of a sequence $(s_n)_{n\in \N}$ of simple functions,
and such that $\intMplus \norm{f - s_n} \,d\mu
\underset{n \rightarrow \infty}{\longrightarrow} 0$, with $\intMplus$ the
Lebesgue integral over {\nonnegative} measurable functions.  In this case it
may be proved that the sequence $\left( \int s_n \, d\mu \right)_{n\in\N}$
is a Cauchy sequence, thus converges, thanks to the completeness of $E$, to a
vector of $E$ that we may define as the integral of $f$.

Therefore, we formally define Bochner-integrable functions as follows.
\begin{Definition}
  A function $f : X \rightarrow E$ is said \emph{Bochner-integrable (with
    regard to $\mu$)} when there exists a sequence $(s_n)_{n\in \N}$ of
  integrable simple functions such that
  \begin{itemize}
  \item
    $\forall x \in X,\; s_n(x)
    \underset{n \rightarrow \infty}{\longrightarrow} f(x)$;
  \item
    $\intMplus \norm{f - s_n} \,d\mu
    \underset{n \rightarrow \infty}{\longrightarrow} 0$.
  \end{itemize}
\end{Definition}
This becomes in {\Coq}
\begin{lstlisting}
Record Bif {f : X -> E} := mk_Bif {
  seq : nat -> simpl_fun E gen;
  ax_notempty : inhabited X;
  ax_int : \forall n, integrable_sf \mu (seq n);
  ax_lim_pw : \forall x : X, is_lim_seq (\fun n => seq n x) (f x);
  ax_lim_l1 : is_LimSup_seq' (\fun n => LInt_p \mu \| f - seq n \|) 0
}.
\end{lstlisting}
A function is therefore Bochner-integrable when there exists such a {\Record}
with the required values and proofs.

Once again, this {\Record} means that in order to prove that $f$ is a
Bochner-integrable function, we need to provide a sequence
\coqe{seq : nat -> simpl_fun E gen} of simple functions, and several properties
corresponding to the mathematical requirements.  In the above {\Record}, we
used \coqe{is_LimSup_seq'} which is a generalization to $\Rbar$-valued
sequences of \coqe{is_LimSup_seq} from {\Coquelicot} that only takes reals.

The hypothesis \coqe{ax_notempty} is artificial.  It is due to our will to be
equivalent to the Lebesgue integral~\cite{BCF21} that requires a {\nonempty}
set for preventing empty lists.  A solution would be to convince the authors of
\cite{BCF21} to switch to our simple functions.

We then prove several lemmas.  First, a Bochner-integrable function is
measurable as it is the pointwise limit of a sequence of measurable simple
functions.  Then, we also prove that $\norm{f}$ is integrable in the sense of
Lebesgue integration of Section~\ref{sec:Lebesgue}.

We also define approximating sequences of integrable simple functions for the
sum, opposite, subtraction, scalar product and norm of a Bochner-integrable
function.  From these proofs and as before, we define useful notations:
\begin{lstlisting}
"bf + bg" := Bif_plus bf bg.
"- bf" := Bif_scal (opp one) bf.
"bf - bg" := Bif_plus bf (Bif_scal (opp one) bg).
"a \cdot bf" := Bif_scal a bf.
"\| bf \|" := Bif_norm bf.
\end{lstlisting}

But such a definition of integrability for vector-valued functions, though easy
to use, does not make it really easy to prove that a given function is
integrable so we look for equivalent properties.  First of all, notice
that if a function $f : X \rightarrow E$ is the pointwise limit of simple
functions, say $(s_n)_{n\in \N}$, then since every $s_n$ has a finite image,
the range of $f$ must be weakly separable according to the previous definition.
And since $f$ must be integrable, we also know that
$\intMplus \norm{f} \, d\mu$ is finite.  Reciprocally, given those two
properties, one may wonder if it is possible to prove that $f$ is
Bochner-integrable.  The answer is yes, and we may even construct an explicit
sequence of simple functions, which is useful for {\Coq} to compute the
value of the integral of~$f$.  To construct such a sequence, we
followed~\cite{tes:tra:21}.  Note that this requires some attention because it
involves some careful splitting of the range of $f$.
\begin{lstlisting}
Lemma Bif_separable_range {f : X -> E} {u : nat -> E} :
  inhabited X -> measurable_fun gen open f -> NM_seq_separable_weak u (inRange f) ->
  is_finite (LInt_p \mu (\fun x : X => \| f \| x)) -> Bif \mu f.
\end{lstlisting}

The first consequence of this characterization is that any measurable function
$f : X \rightarrow \R$ such that $\intMplus \norm{f} \, d\mu < \infty$ is
Bochner-integrable, because we already know that $\R$ is (weakly) separable.
\begin{lstlisting}
Lemma R_Bif {f : X -> R_NormedModule} :
  inhabited X -> measurable_fun gen open f -> is_finite (LInt_p \mu (\| f \|)) -> Bif \mu f.
\end{lstlisting}

\medskip

So here we recover exactly the definition of integrability for the Lebesgue
integral, which makes both definitions compatible.  But now, if we assume $X$
separable and $f$ continuous, one can prove that the range of $f$ is separable.
For example, we deduce that every continuous function $f : \R^n \rightarrow E$
is Bochner-integrable.

\medskip

For other cases where our function $f$ seems too complicated to prove easily
that its range is separable, we still have the ability to prove its Bochner
integrability by using yet another equivalent property.

\begin{Lemma}
  A function $f : X \rightarrow E$ is Bochner-integrable if and only if
  \begin{itemize}
  \item
    it is the pointwise limit of a sequence of simple functions (without
    requiring any integrability);
  \item
    $\intMplus \norm{f} \, d\mu < \infty$.
  \end{itemize}
\end{Lemma}

Functions which are pointwise limit of simple ones ({\ie} which satisfies the
first dot above) are said \emph{strongly measurable}.  This definition has been
formalized inside the library too, and we have proved several useful properties
about strongly measurable functions.  The most striking example is that any
pointwise limit of strongly measurable functions is still strongly measurable.
Such a property is of great use to prove the dominated convergence theorem.

\section{The Bochner integral}
\label{sec:integral}

The definition of the Bochner integral is straightforward from the
integrability definition.  Let us still consider a measure space
$(X,\Sigma,\mu)$, and~$E$ a Banach space over~$\R$.
\begin{Definition}
  Let $f : X \rightarrow E$.
  Let $(s_n)_{n\in \N}$ be a sequence of integrable simple functions such that
  \begin{itemize}
  \item
    $\forall x \in X,\; s_n(x)
    \underset{n \rightarrow \infty}{\longrightarrow} f(x)$;
  \item
    $\intMplus \norm{f - s_n} \,d\mu
    \underset{n \rightarrow \infty}{\longrightarrow} 0$.
  \end{itemize}
  Then the \emph{Bochner integral of $f$ (relatively to measure $\mu$ on $X$)}
  is defined by
  $$\int f \, d\mu := \lim_{n \to \infty} \int s_n \, d\mu.$$
\end{Definition}

The {\Coq} definition is quite short because the partial function
\coqe{lim_seq} was used, so the convergence of the sequence do need to be
checked while defining \coqe{BInt} However, it was proved as an independent
lemma, which is essential to be able to use the properties of \coqe{BInt bf} as
a limit.
\begin{lstlisting}
Definition BInt {f : X -> E} (bf : Bif \mu f) := lim_seq (\fun n => BInt_sf \mu (seq bf n)).
\end{lstlisting}

The first property to ensure is that this definition does neither depend on the
chosen sequence $(s_n)_{n\in\N}$, nor on the integrability proof.  This is
stated as the following extensionality lemma.
\begin{lstlisting}
Lemma BInt_ext {f f' : X -> E} :
  \forall (bf : Bif \mu f) (bf' : Bif \mu f'), (\forall x : X, f x = f' x) -> BInt bf = BInt bf'.
\end{lstlisting}

We then prove all the expected properties of the integral such as linearity or
the triangular inequality, by taking the limit of the already proved properties
over simple functions.

A larger proof is the equality of \coqe{BInt} and \coqe{LInt_p} for
{\nonnegative} real-valued integrable functions, as we had to prove the
equivalence of the two formalizations (of simple functions, of integrability
and of integrals).  It makes our library compatible with the one about the
Lebesgue integral.

The next lemmas were chosen to ease the main perspective of this work, that is
the definition of Bochner spaces in {\Coq}, which are a generalization of $L^p$
spaces for the Lebesgue integral.

\begin{Theorem}
  A function $f : X \rightarrow E$ is zero $\mu$-almost everywhere if and only
  if $\intMplus \norm{f} \, d\mu = 0$.
\end{Theorem}

\begin{Theorem}[dominated convergence]
  Given a {\nonnegative} integrable function $g : X \rightarrow \R$, and a
  pointwise convergent sequence $(f_n)_{n\in\N}$ of Bochner-integrable
  functions such that $\forall n \in \N$, we have $\norm{f_n} \le g$, then
  $f := (x \mapsto \lim_{n \to \infty} f_n(x))$ is Bochner-integrable, and
  $$\lim_{n \to \infty} \int f_n \,d\mu = \int f \,d\mu.$$
\end{Theorem}

\section{Conclusion and perspective}
\label{sec:concl}

We have defined the Bochner integral with a constructive point of view for
Bochner integrability.  We have proved that a function is Bochner-integrable
(with the constructive dependent type definition) provided it is the pointwise
limit of simple functions and that its range is weakly separable.  We have also
proved that our definitions are consistent with those of a Coq formalization of
the Lebesgue integral.

Our design choices are twofold.  Mathematically, we have conscientiously
followed Teschl~\cite{tes:tra:21} with a kind of \emph{weak} separability
instead of (regular) separability.  Formally, we have simple functions with an
index function and Bochner-integrable by a dependent type.  We have succeeded
in proving the common lemmas, from linearity to dominated convergence so this
seems a good basis to build upon.

\medskip

This opens the way to the formalization of Bochner spaces of strongly
measurable functions for which the $p$-th power of the norm is Lebesgue
integrable.  They are a generalization of the usual $L^p$ Lebesgue spaces,
where functions equal almost everywhere are also identified.  Such spaces are
also Banach spaces for $p\geq1$.  For instance, given a regular enough space
domain $\Omega\subset\R^3$, the square-integrable functions~$\Omega\to\R$ form
the Hilbert space~$L^2(\Omega)$ since~$\R$ is a Banach space on which Bochner
integration applies.  Moreover, given a real number~$T>0$, the
square-integrable functions from~$[0,T]$ to~$L^2(\Omega)$ also form the Hilbert
space $L^2([0,T],L^2(\Omega))$.  And eventually, this could be used to apply
the Lax--Milgram theorem~\cite{BCF17} in the context of the resolution of some
set of partial differential equations.

\bibliographystyle{plainnat}
\bibliography{biblio}

\end{document}